\documentclass[11pt]{article}
\usepackage{hyperref}
\pdfoutput=1

\begin{document}
\title{Optically induced electrokinetic patterning \\ and manipulation of particles}
\author{Stuart J. Williams, Aloke Kumar, Steven T. Wereley \\
\\\vspace{6pt} School of Mechanical Engineering,
\\ Purdue University, West Lafayette, IN 47907, USA}
\maketitle
%% The abstract (in this file, and that submitted as text to arXiv) should include the exact phrase
%% "fluid dynamics video" or "fluid dynamics videos"
\begin{abstract}
This fluid dynamics video showcases how optically induced electrokinetic forces can be used to drive three-dimensional micro-vortices. The strong microfluidic vortices are used constructively in conjunction with other electrokinetic forces to dynamically and rapidly aggregate particle groups. Particle manipulation is achieved on the surface of a parallel-plate gold/indium tin oxide (ITO) electrode that is illuminated with near-infrared (1064 nm) optical patterns and biased with an low frequency ($<$ 100 kHz) alternating current (AC) signal. The fluid dynamics video shows how electrokinetically driven flows in the microdomain can be used for non-invasive particle manipulation.
\end{abstract}
% main text
\section{Introduction}

%% The format is: \href{URL of video}{name that will appear in the text}
The sample video is
\href{http://ecommons.library.cornell.edu/bitstream/1813/11399/2/REPmovie-mpg1.mpg}{Video1} \\
\\

The rich phenomenology of electrokinetically driven flows has produced extremely diverse applications of fluid flow for `lab-on-a-chip' type of applications. Amongst these, the ability to capture, manipulate and assemble colloidal particles and biological cells is of importance for a wide range of micro-engineering applications. Here we investigate a new method of non-invasive fluid and particle manipulation technique that utilizes optically induced AC electrokinetic mechanisms. Our method utilizes near-infrared (1064 nm) light patterns generated from a holographic illumination system (Bioryx 200, Arryx Inc, Chicago, USA) coupled with a simple parallel electrode plate micro-fluidic platform with two different materials (ITO and gold) deposited on glass substrates. The simultaneous application of highly-focused light and an AC electric field will induce fluid flow towards the illuminated region and particles are subsequently accumulated on the electrode surface. The accumulated particles can be readily assembled, translated and patterned anywhere on the surface of the gold substrate. By changing the shape and intensity of the optical landscape unique geometrical colloidal assemblies can be dynamically configured. The particle group can be translated by changing the location of the optical illumination, physically moving the stage, or deactivating the initial illumination and activating a separate source at a different location. Optically induced electrohydrodynamics and electrokinetic forces are responsible for this particle manipulation technique. 
\\
\\
The focused illumination in the parallel plate electrode platform generates a unique toroidal microvortex. Due to the uniform electric field from the parallel-plate design, AC electro-osmosis (fluid motion subjected to non-uniform electric fields at low frequencies) was not observed. This microfluidic vortex shows promise to investigate electro-thermal flows decoupled from AC electro-osmosis and other electrokinetic phenomena. 
\\
\\
In this video, fluorescent polystyrene particles of 1.0 micrometers are manipulated in DI water with an illumination intensity of no more than 20 mW. The parallel electrode plates are separated by a distance of 50 micrometers. The applied electric field can be variable; the AC frequency was varied in between 10 to 50 kHz , while the applied potential was kept under 20 volts peak-to-peak.
\\
\\

\section{References}
\begin{enumerate}
\item
S.J. Williams, A. Kumar, and S.T. Wereley (2008). "Electrokinetic patterning of colloidal particles with optical landscapes, doi: 10.1039/b810787d." Lab on a Chip (2008).
\item
 A. Kumar, S. J. Williams, and S.T. Wereley (2008). "Experiments on opto-electrically generated microfluidic vortices, doi: 10.1007/s10404-008-0339-8." Microfluidics and Nanofluidics (2008).
\item
 S.J. Williams, A. Kumar, S.D. Peterson, H-S. Chuang, and S.T. Wereley, "Three-dimensional transport of an optically induced electrothermal microvortex," Amer. Phys. Soc./Div. Fluid Dyn. Annual Meeting, San Antonio, TX (2008).

\end{enumerate}
\end{document}